\documentclass{article}

\usepackage[left=0.5in, right=0.5in, top=1in, bottom=1in]{geometry}
\usepackage[colorlinks]{hyperref}
\usepackage{amsmath}
\usepackage{amssymb}
\usepackage{newpxtext}
\usepackage{newpxmath}
\usepackage{orcidlink}
\usepackage{biblatex}
\usepackage{graphicx}
\usepackage[font=small]{caption}

\definecolor{KUred}{RGB}{144, 26, 30}
\definecolor{randomblue}{HTML}{547AA5}

\hypersetup{citecolor=blue, linkcolor=KUred, urlcolor=black}

\DeclareCaptionLabelSeparator{colon}{\textcolor{KUred}{\textbf{{:\space}}}}
\DeclareCaptionLabelFormat{redbold}{\textcolor{KUred}{\textbf{#1 #2}}}
\newcommand{\sgn}{\mathrm{sgn}}

\begin{document}

\title{Delayed control driven oscillations in plant roots}

\author{
Riz Fernando Noronha \orcidlink{0009-0007-2923-3835}$^{1\,\dagger}$, 
Kunihiko Kaneko \orcidlink{0000-0001-6400-8587}$^{1}$,
Koichi Fujimoto \orcidlink{0000-0001-6473-7990}$^{2}$
}

\date{$^{1}$\textit{Niels Bohr Institute, University of Copenhagen, Copenhagen, Denmark}\\ \vspace{5pt}
$^{2}$\textit{Program of Mathematical and Life Sciences, Hiroshima University, Higashi-Hiroshima, Japan}\\ \vspace{10pt}
$^\dagger$ Corresponding author: \href{noronha@nbi.ku.dk}{noronha@nbi.ku.dk}}

\maketitle

\renewcommand{\abstractname}{\large Abstract\\}

\begin{abstract}
    \normalsize
    \textit{Arabidopsis} roots show oscillatory growth patterns on homogeneous agar surfaces, whereas other plants, such as maize, do not. Although several explanations have been proposed, a simple and general model that makes testable predictions across species has been lacking. Roots sense gravity and correct their growth direction towards the vertical. Motivated by recent evidence for a time delay in this gravitropic correction, we develop a minimal nonlinear model based on the delay hypothesis that predicts whether a root oscillates or grows vertically downwards. The model identifies a fourfold relation between the delay and time period, robust across different response functions. Analysing images of \textit{Arabidopsis}, we find that the mode of the oscillatory arc length is not significantly different between inclined and vertical growth conditions. The quantitative agreement between the experimentally measured oscillatory arc length and the arc length estimated from estimated root growth speed and response delay supports this fourfold delay–period rule for delay-driven root oscillations. The simplicity of our model allows for a direct comparison with data from diverse plant species.
\end{abstract}

\twocolumn

\section{Introduction}

Plant roots are known to exhibit complex patterns in soil. As real soil is far too difficult to understand, experiments on root growth are typically performed in a homogeneous environment of agar surface. The model species \textit{Arabidopsis} show a particular type growth: a single root takes a `waving' pattern, especially when grown on an incline \cite{okada1990reversible}. This phenomena is reproducible among genetically identical clones of the same plant, however different species do not show such oscillations. Despite several proposed explanations, there remains a lack of a simple yet universal model that can make predictions across species with only a few parameters.

The helical growth of plants (termed circumnutation) \cite{agostinelli2020nutations} was first studied by Darwin \cite{darwin1880power}, who proposed that an internal oscillator could be the cause. However, it is not sufficient to explain large deviations: if a root is placed at a horizontal position, it bends and returns to the vertical without any oscillatory motion, as would be expected from an internal oscillator.

Gradmann \cite{gradmann1921uberkrummungsbewegungen} proposed an alternate theory of overshooting, where a gravitational response is sufficiently large to cross over in the other direction. Israelsson and Johnson \cite{israelsson1967theory} provided a theory for the following, in a simple and elegant 1-dimensional model, which described the gravitropic response as a delayed control. Delay differential equations like the one they propose are often studied in physics \cite{balanov2004controldelayedfeedback, janson2004delayed, mackey1977oscillation}, notably in the Ikeda model which exhibits a variety of nonlinear dynamics including chaos \cite{ikeda1982successive, ikeda1987high}.

Recently, Porat et al. \cite{porat2024mechanical} propose a mechanical model that suggests twisting is a major factor in oscillations in plant roots. While twisting is undoubtedly an important factor \cite{thompson2004root}, especially when grown on highly inclined surfaces, other experimental results, such as that of Buer et al \cite{buer2003ethylene} suggest that waving is possible even in the absence of twisting.

There exists a small but growing body of evidence that shows the presence of a time delay in the gravitropic response of \textit{Arabidopsis} roots. The current explanation is that root bending occurs due to differential cell growth: one side grows faster than the other, leading to curvature. This differential cell growth is believed to be modulated by auxin. Experiments performed with a root starting at the horizontal indicate that although the timescale of the gravity sensing response amyloplasts in the root tip is very fast (around a minute), there is a time lag of approximately 2 hours before the auxin gradient between the top and the bottom is established \cite{band2012root, baldwin2013gravity}. In other words, once a root senses that its orientation is incorrect, there is a time-delay of 2 hours before it appropriately responds to correct its path.

\section{Model}

We present a first order stochastic delay differential equation to describe the angle from the vertical of the root tip, $\theta$:

\begin{equation} \label{main-equation-sdde}
    \frac{\textrm{d}\theta}{\textrm{d}t} = \underbrace{- k \,\sin(\theta(t-\tau))}_{\text{Gravitropy}} + \underbrace{\eta\, \xi(t)}_{\text{Noise}}
\end{equation}

The parameter $k$ represents the sensitivity to gravity: the negative sign implies negative feedback, as $\theta>0$ implies $\dot{\theta}<0$, and vice versa. Biologically, it can be thought of the strength of the gravitropic response triggered by a maximum gravitropic stimulus (in this case, assumed to be the root at an angle of 90 degrees from the vertical, following previous data \cite{mullen2000kinetics}). It can thus be linked to the magnitude of the differential auxin gradient established across the top and bottom of the root.

We introduce a time-delay of $\tau$ into the gravity sensing. As discussed earlier, the source of this time delay comes from the $\approx$2 hours required for the appropriate auxin gradient to be established, following a gravity stimulus \cite{band2012root, baldwin2013gravity}. We choose to use a sinusoidal non-linearity in the response, as in addition to being a simple, it was previously discussed emperically from data \cite{mullen2000kinetics}.

As root growth is a noisy process even in agar, we use additive Gaussian white noise $\xi(t)$, with magnitude $\eta$. The model resembles the linear SDDE (Stochastic Delay Differential Equation) previously studied \cite{ohira2000delayed, kuchler1992langevins}, with the addition of a sinusoidal nonlinearity. While other non-linear SDDEs have been studied \cite{ando2017time, albers2022chaotic}, this simple non-linear model remains relatively unexplored. The deterministic case ($\eta=0$), introduced by Wischert et al. \cite{wischert1994delay} has been well-studied by Schanz and Pelster \cite{schanz2003analytical} and Sprott \cite{sprott2007simple}.

Note that mathematically, the three parameters $\tau, k, \eta$ can be reduced into a simple 2-parameter system by rescaling the time $t$. However, we adopt the more complicated notation as they biologically correspond to different effects.

\section{Results}

\subsection{Deterministic DDE}

In the zero-noise case ($\eta=0$) we can, for small $\theta$, approximately linearize \autoref{main-equation-sdde} to be
\begin{equation} \label{noiseless-linear-dde}
    \frac{\textrm{d}\theta}{\textrm{d}t} = - k \,\theta(t-\tau)
\end{equation}
This can be solved to give the solution (see \autoref{appendix:linear-model})
\begin{equation} \label{noiseless-linear-soln}
    \theta(t) = A\, e^{\lambda t}
\end{equation}
Where $\lambda$ is the solution to the characteristic equation
\begin{equation} \label{noiseless-linear-characteristic-eqn}
    \lambda + k\,e^{-\lambda \tau} = 0
\end{equation}
$\lambda$ is, in principle, complex, and can be written as $\lambda= a + i\,\omega$. The stability depends on the value of $a$: $a<0$ leads to exponential damping, while $a>0$ is unstable.

We can solve \autoref{noiseless-linear-characteristic-eqn} at the critical point where $\lambda = i\,\omega$ (see \autoref{appendix:linear-model}), and derive
\begin{equation} \label{critical-boundary}
    k\,\tau = \frac\pi2
\end{equation}
which is the equation of the critical boundary.

\begin{itemize}
    \item[-] $k\,\tau < \pi/2$ leads to damped oscillations.
    \item[-] $k\,\tau = \pi/2$ is the critical value, where oscillations are neither damped nor enhanced.
    \item[-] $k\,\tau > \pi/2$ is unstable, with growing oscillations.
\end{itemize}

\begin{figure}
    \centering
    \includegraphics[width=\linewidth]{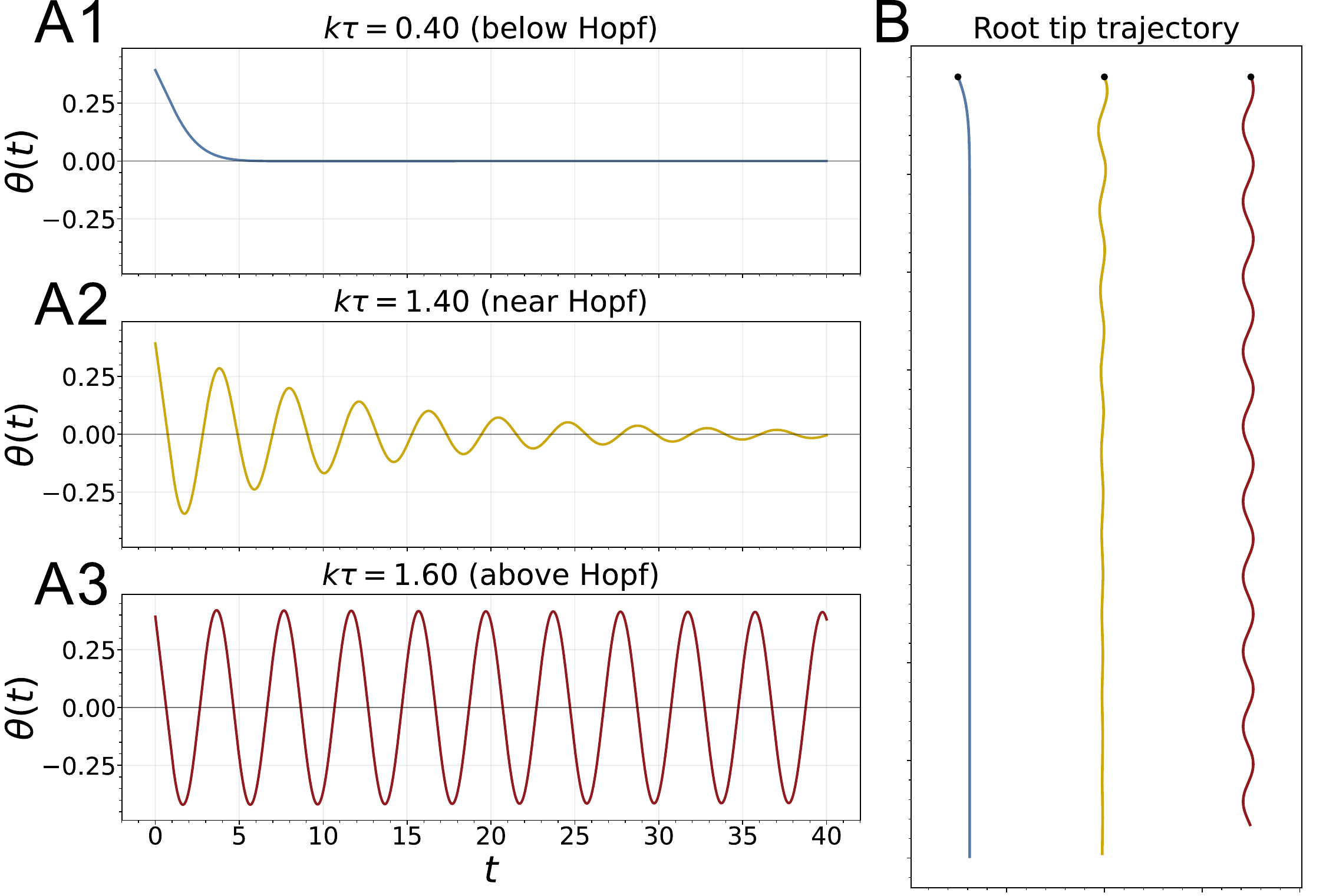}
    \caption{An illustration of how crossing the Hopf bifurcation can lead to oscillatory dynamics, for the deterministic ($\eta=0$) case of \autoref{main-equation-sdde}. The inital function is chosen to be $\theta(t) = -\pi/8$ for all $-\tau \leq t \leq 0$. A1: Far below the Hopf bifurcation, the system behaves as if it would with no time delay $\tau$, and decrease to $\theta=0$. A2: Closer to the Hopf bifurcation, the system still decays, but damped oscillations are visible. A3: Above the Hopf bifurcation, a stable oscillatory solution is acheived (note that in the linear system described by \autoref{noiseless-linear-dde}, the oscillations increase unboundedly, but the non-linearity of $\sin$ allows for a stable oscillatory behavior to emerge). B: A trace of root tip trajectories following the dynamics in A.}
    \label{fig:hopf-bifurcation-simple}
\end{figure}

As the eigenvalues cross over the imaginary axis, the emergence of oscillations is due to a Hopf bifurcation, in particular a delay-induced Hopf bifurcation (as without the delay, $\tau=0$, the system would always be damped and return to the vertical).

The parameters for $k$, the strength of the gravitropic response, and $\tau$, the time-lag in gravity sensing, cannot be easily identified. However, at the critical point, we can calculate that $\omega_0 = k$. As $\omega_0$ represents the frequency of the oscillations, the time period can be calculated to be

\begin{align}
    T &= \frac{2\,\pi}{k} \\
    &= 4\, \tau \label{factor-4-equation}
\end{align}

Above the critical point, the linear system diverges. In order to study the oscillations, we need to consider a non-linearity that allows for stable limit cycles to form. We choose to use $\sin(x)$ as in \autoref{main-equation-sdde}, as a simple approximation.

Increasing $k$ in the noise-less model is observed to lead to chaos. Although our system is continuous, we can stroboscopically discretize it in intervals of $\tau$ in order to draw an analogy to a dynamical map. In that case, the zero-noise case resembles the sine map: $x_{n+1} = k \sin(x_n)$ \cite{Arnold1961Small}, or perhaps more closely the circle map with no externally applied frequency $x_{n+1} = x_n + k \sin(x_n)$ \cite{jensen1983complete, kaneko1986collapse} which also show chaotic behaviour. Both the sine and circle maps approach chaos through the period doubling route \cite{feigenbaum1978quantitative, feigenbaum1983universal, grossmanPeriodDoubling, Myrberg1963}: The system goes from a period of 1, to 2, to 4, to 8, and so on until it devolves into chaos. However, there's a significant difference in our delay-induced oscillations. The period of the primary limit cycle's oscillations is typically fixed at $4\tau$, as the oscillations are formed due to the time delay. Thus, $k$ cannot typically increase the time period (as the time period is only dependent on $\tau$), yet for a large $k$, we would expect chaotic behaviour: the update causes extremely strong deviations, which are then normalized by the sine function (which maps even very large arguments to the interval $\left[0,1\right]$). Numerical simulations agree with this: we do observe chaotic behavior for large $k$, but the route to chaos is not through period doubling, but through intermittency \cite{pomeau1980intermittent}.

\begin{figure}
    \centering
    \includegraphics[width=\linewidth]{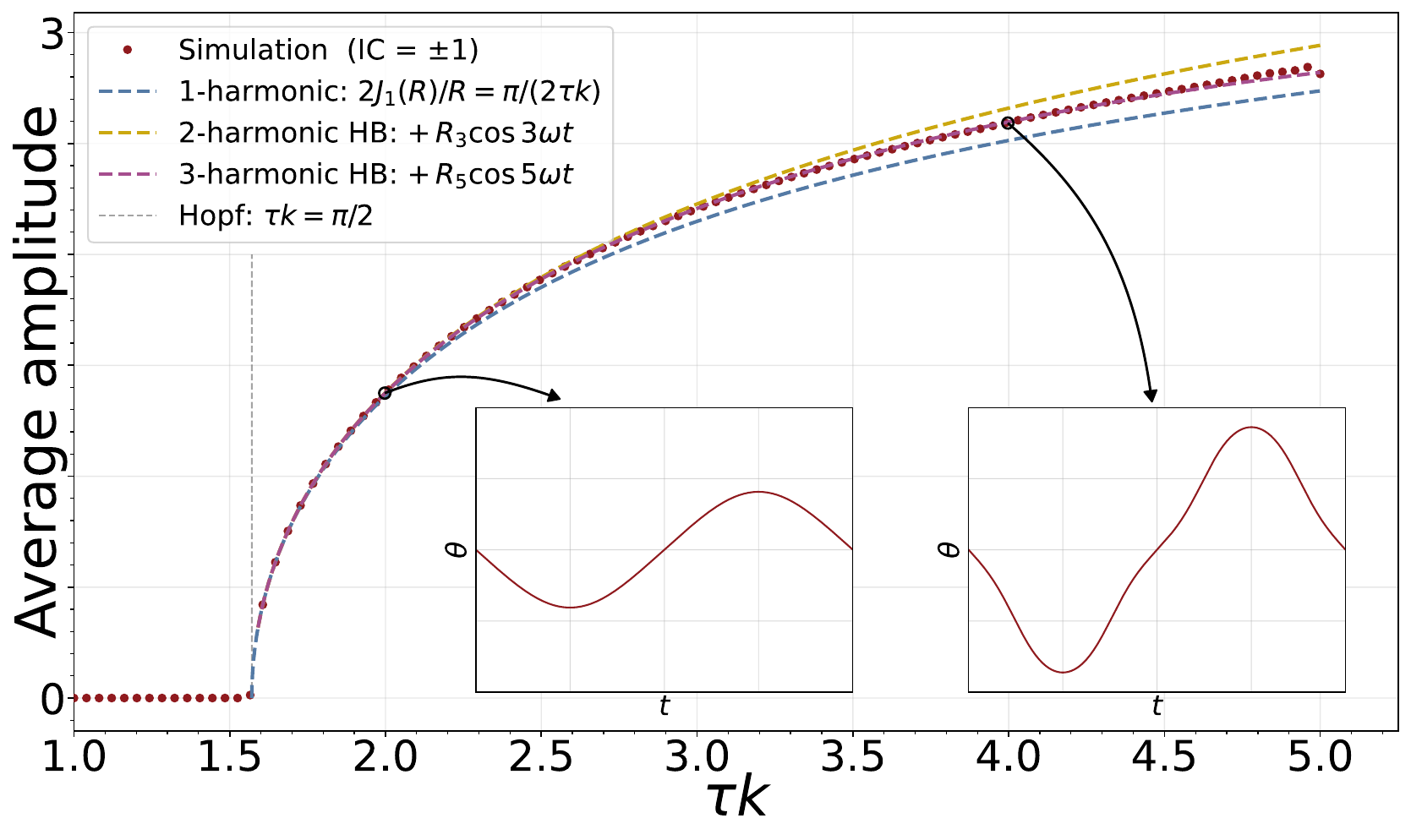}
    \caption{The amplitude of the period 4 limit cycle (red) versus the control parameter $\tau k$. Dashed lines represent the analytical approximations made through the Jacobi-Anger expansion. The first order (blue) line described by \autoref{amplitude-scaling-first-order} is a good fit initially, but breaks down at larger $\tau k$ due to the oscillations not being well-approximated as sinusoids, as shown in the insets figures. Numerically, using 3 terms (magenta) in the expansion is enough to approximate the behavior reasonably well.}
    \label{fig:amplitude-vs-tauK}
\end{figure}

The amplitude of the 4-period oscillation is noted to increase with an increase in $k\,\tau$. By assuming a sinusoidal solution of the form $\theta(t) = R\cos(\omega t)$, and using the Jacobi-Anger expansion, we can derive (see \autoref{appendix:amplitude-scaling})
\begin{equation} \label{amplitude-scaling-first-order}
    k\,\tau = \frac{\pi R}{4 J_1 (R)}
\end{equation}
Where $R$ is the amplitude of the oscillations, and $J_1$ is the first Bessel function of the first kind. The theoretical amplitude agrees somewhat well with numerical results, though as $k\,\tau$ increases, higher harmonics become more prevalent. By assuming a $n$ harmonic ansatz instead, and expanding the Jacobi-Anger expansion up to $n$ terms, we are able to make higher-order theoretical predictions. A comparison to the numerical data shows that we obtain a reasonable approximation at $n=3$ (\autoref{fig:amplitude-vs-tauK}).

\subsection{Generalized response functions}

We can write a more general case with a delayed response as
\begin{equation} \label{delayed-control-general-f}
    \frac{\textrm{d}\theta}{\textrm{d}t} = - k f(\theta(t-\tau)) + \eta\, \xi(t)
\end{equation}
Where $f$ is the control function. Assuming that $f$ can be linearized around $0$, we can expect a Hopf bifurcation as previously discussed. After the Hopf bifurcation, we can construct a heuristic argument as to why, if limit cycles exist, the period should be $\approx 4$. If $\theta(t_1-\tau) < 0$, the `restoring force' wants to increase $\theta(t_1)$, regardless of the current value. Consider the case where $\theta(t_1-\tau) = 0$, just crossing the origin. In that case, $\dot{\theta}(t_1)=0$, which means we should expect a maxima. After that point, $\theta(t_1 - \tau + \Delta t)$ is positive, which means that the function should decrease, until it once again reaches the point $\theta(t_2)$ where $\theta(t_2 - \tau)=0$, leading to a minimum. From this, we can predict that the distance from the x-intercept to the crest/trough is $\tau$: Assuming that the limit cycle is approximately symmetrical allows us to express the time period of the oscillations as $4\tau$.

It is worth noting that the presence of the 4-period limit cycle seems somewhat universal: we speculate that any function $f$ which satisfies the following criteria will usually show oscillations of period $T\approx 4 \tau$ in the delayed control case of \autoref{delayed-control-general-f}:

\begin{enumerate}
    \item $f$ can be linearized around $x=0$, showing a Hopf bifurcation at $\approx \pi/2$,
    \item \label{conditions:item2} $f$ is an odd function (so the response is symmetric about the origin),
    \item $f(x)> 0$ for all $x>0$, and thus (by \ref{conditions:item2}) $f(x) < 0$ for all $x < 0$
    \item After the Hopf bifurcation, $f$ shows stable limit cycles and does not diverge.
\end{enumerate}

\begin{figure}
    \centering
    \includegraphics[width=\linewidth]{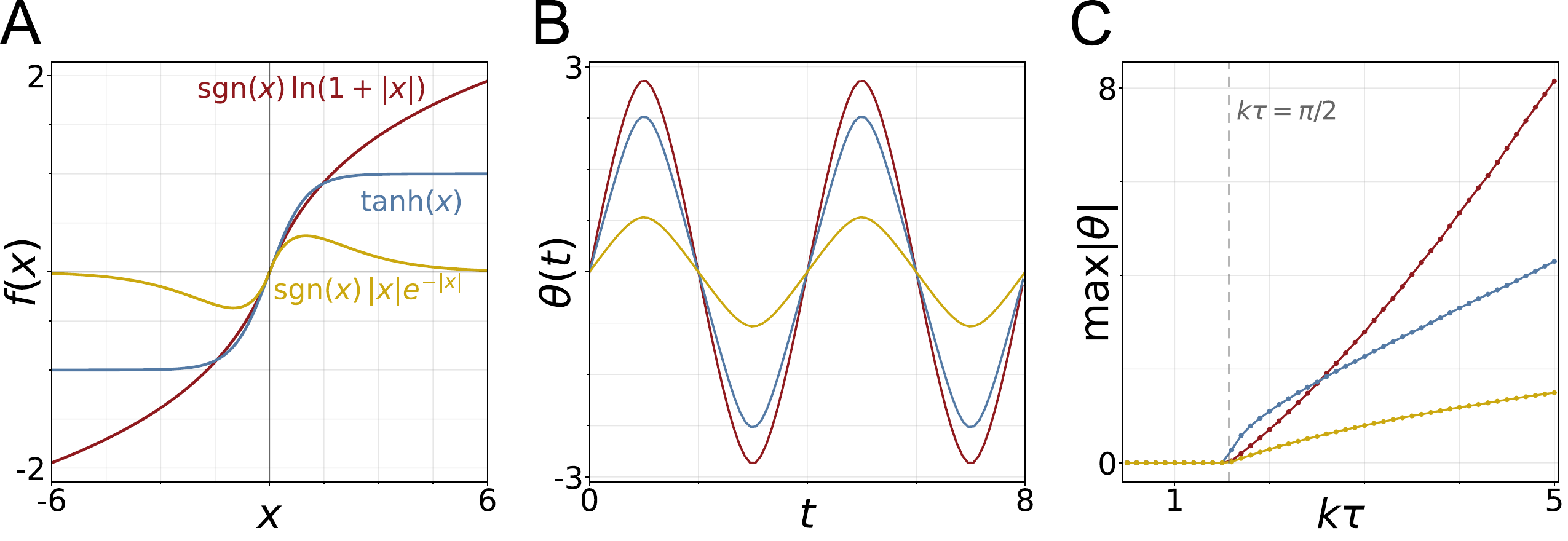}
    \caption{A comparison of the oscillatory behavior observed for non-sinusoidal response functions $f(x)$, for the DDE described in equation \autoref{delayed-control-general-f}. A: The three test functions. B: The attractor for $\tau=1, k=3$ (calculated for initial functions $\theta(t)=1$ for all $-\tau\leq t\leq0$). Limit cycles of period 4 appear to be universal. C: The amplitude scaling of the test functions.}
    \label{fig:other-response-functions}
\end{figure}

To test this, we simulate three other non-linear functions $f$. As $f(x)=x$ diverges and is unstable, we tried $f(x)=\tanh(x)$ (which is asymptotic to the line $y=1$), $f(x)=\sgn(x) \ln(\lvert x \rvert + 1)$ (which grows slower than the linear case), and $f(x)=\sgn(x)\lvert x\rvert e^{-\lvert x\rvert}$ (which reaches an optimum and decreases, ultimately being asymptotic to $y=0$). All three cases were observed to form limit cycles of period $T\approx4$ after the Hopf bifurcation (although there were other differences, such as in the amplitude scaling with $\tau\,k$), as shown in \autoref{fig:other-response-functions}.

\subsection{Stochastic DDE}

We now turn our attention back to the sinusoidal mapping $f(x)=\sin(x)$ described in the deterministic case of \autoref{main-equation-sdde}. As previously observed \cite{schanz2003analytical}, at $k\tau$, a bifurcation causes a second limit cycle with period $> 4$ to come into existence. This may seem puzzling as it does not align with our previous argument for period$=4$. However the argument relied on the assumption that $f(x) = 0$ only at $x=0$, which is not the case for $\sin(x)$, which is also $0$ at $\pi, 3\pi,$, etc. As such, if a limit cycle of amplitude $R>\pi$ were to exist, as it crosses $\pi$, it would seed two different optima (where $d\theta/dt=0$) at time $\tau$ in the future. 


\begin{figure}
    \centering
    \includegraphics[width=\linewidth]{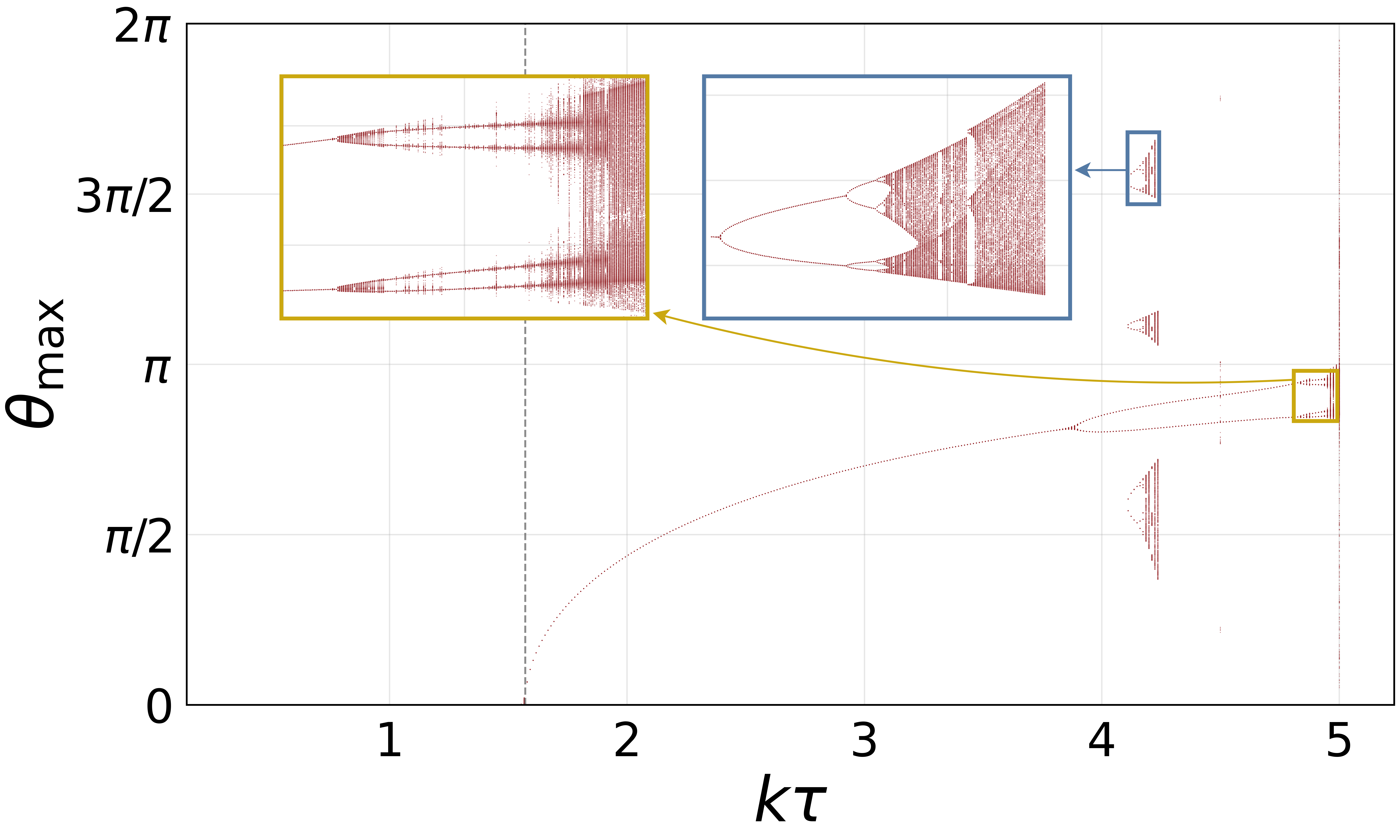}
    \caption{The bifurcation diagram of the deterministic sinusoidal model (\autoref{main-equation-sdde} with $\eta=0$) created by plotting $\theta_\mathrm{max}$, the value of $\theta$ where $\mathrm{d}\theta/\mathrm{d}t=0$ and $\mathrm{d}^2\theta/\mathrm{d}t^2 < 0$, showing the appearance of the second limit cycle, and the approach to chaos.}
    \label{fig:bifurcation-diagram}
\end{figure}

The second limit cycle of the sinusoidal mapping appears to arise due to a saddle node of limit cycles, also known as a fold bifurcation. This can be seen by the scaling of transients of the second limit cycle before the bifurcation. The transients have finite lifetimes, and through the normal form of the saddle node bifurcation, we can predict that the lifetimes scale as a power law with critical exponent $1/2$, matching data from numerical simulations (\autoref{appendix:ghost-escape-times}). As shown in the bifurcation diagram (\autoref{fig:bifurcation-diagram}), the volume of the chaotic attractor is expanded before it suddenly disappears: which suggests the bifurcation known as a crisis (where the chaotic attractor collides with an unstable periodic orbit) \cite{grebogi1986critical}. Here, the critical exponent is numerically identified to be $\approx 0.68$, distinct from the $1/2$ of a saddle-node bifurcation (\autoref{fig:lc2-ghost-escape-times}).

We then study the effect of noise by using the autocorrelation $C(\Delta) = \langle\theta(t)\theta(t-\Delta)\rangle$. In the zero noise case ($\eta$=0), the system is perfectly oscillatory (above the bifurcation) and thus the autocorrelation is also oscillatory with the same period. A non-zero noise introduces phase diffusion to the delay-induced oscillatory state. This phase diffusion should decrease the autocorrelations (as, phases are shifted over long delay time $\Delta$, though the random noise has not got a high chance to affect the phase over small $\Delta$), and thus the autocorrelation function should resemble `damped' oscillations.

However, observations from numerical simulations show a `transition' in the damping rate. Rather than the damping rate increasing smoothly as a function of the noise strength $\eta$ \footnote{This would be the case for a simple harmonic oscillator with noise on the phase term, i.e, $d^2\theta/dt^2 = -\omega_0^2 \sin(\theta + \eta \xi(t))$. A simple harmonic oscillator with additive noise, such as $d^2\theta/dt^2 = -\omega_0^2 \sin(\theta) + \eta \xi(t)$ would not have phase drift, and remain coherent with only its amplitude being affected by the noise strength $\eta$.}, we observe that at low-noises, the damping is relatively constant (but non-zero). Beyond a certain critical threshold, the damping begins to scale with $\eta$.

In order to investigate this transition in the autocorrelation damping, we study the return maps of the system. We simulate trajectories of $\theta$, and (after reaching an oscillatory steady state), look at how $\theta(t+\tau)$ depends on $\theta(t)$ (in other words, we `discretize' our system to compare it to a non-linear stochastic map, with time-steps of $\tau$.

\begin{figure}
    \centering
    \includegraphics[width=\linewidth]{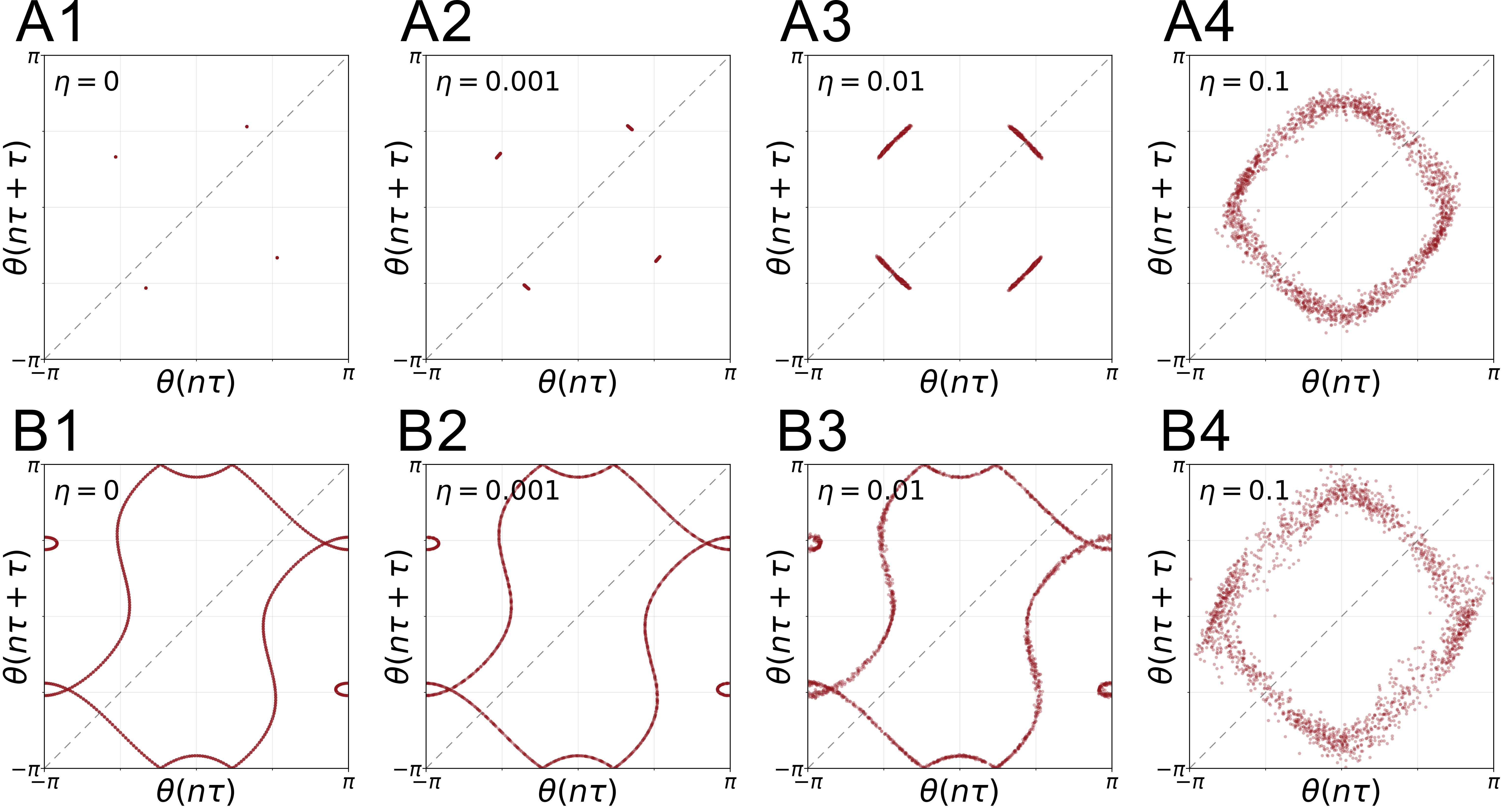}
    \caption{A plot of the stroboscopic return maps $\theta(n\tau + \tau)$ versus $\theta(n\tau)$, from \autoref{main-equation-sdde}. A: $\tau=1, k=3$, showing the period 4-limit cycle. As the noise increases, the points on the return map `diffuse'. However, as long as they're distinct (eg: $\eta=0.001$), noise cannot cause sudden phase shifts, leading to persistent oscillatory autocorrelations. At $\eta=0.01$, a blurred state is achieved where the same value of $\theta(t)$ can map onto different values of $\theta(t+\tau)$, leading to a sudden damping of the autocorrelation. B: $\tau=1, k=4$, showing the second limit cycle. As the periodicity is not $4$, the pattern cannot be easily interpreted: however, at high noise, it degenerates into the noisy period 4-limit cycle.}
    \label{fig:return-maps-noise}
\end{figure}

4-cycles are present in the system, which can be seen by plotting the stroboscopic return map ($\theta(nt + \tau)$ vs $\theta(nt)$, for $n=1,2,\dots$), as in \autoref{fig:return-maps-noise}A. The four points correspond to the four phases of one oscillation, cycling in the order $A\to B\to C \to D \to A$. With weak noise, each point broadens into a small `cloud', but the ordering is preserved: the noisy sets $\tilde{A}, \tilde{B}, \tilde{C}, \tilde{D}$ still map sequentially by $\tilde{A}\to \tilde{B}\to \tilde{C} \to\tilde{D} \to \tilde{A}$. As the clouds remain well separated, the phase cannot jump, and the period of $4\tau$ is preserved. At a critical noise strength, however, the cloud around $\tilde{A}$ grows such that some points in it map to $\tilde{C}$ as well as $\tilde{B}$. This breaks the ordering of the 4‑cycle and allows sudden phase shifts over a single $4\tau$ interval. This allows the phase to dramatically change after $4\tau$ time steps, leading to a sharp change in the autocorrelation damping.

\section{Comparison to Data}

We analyze the waving patterns of the Col-0 wild type accession of \textit{Arabidopsis thaliana}, grown on both inclined and non-inclined surfaces of \textit{Arabidopsis} growth medium \cite{Okada1992} supplemented with 1 $\%$ (w/v) sucrose and 1 $\%$ (w/v) agar under constant illumination. We use a snapshot image for each growing root, and make the assumption that the root growth follows that of the tip.

\begin{figure}
    \centering
    \includegraphics[width=\linewidth]{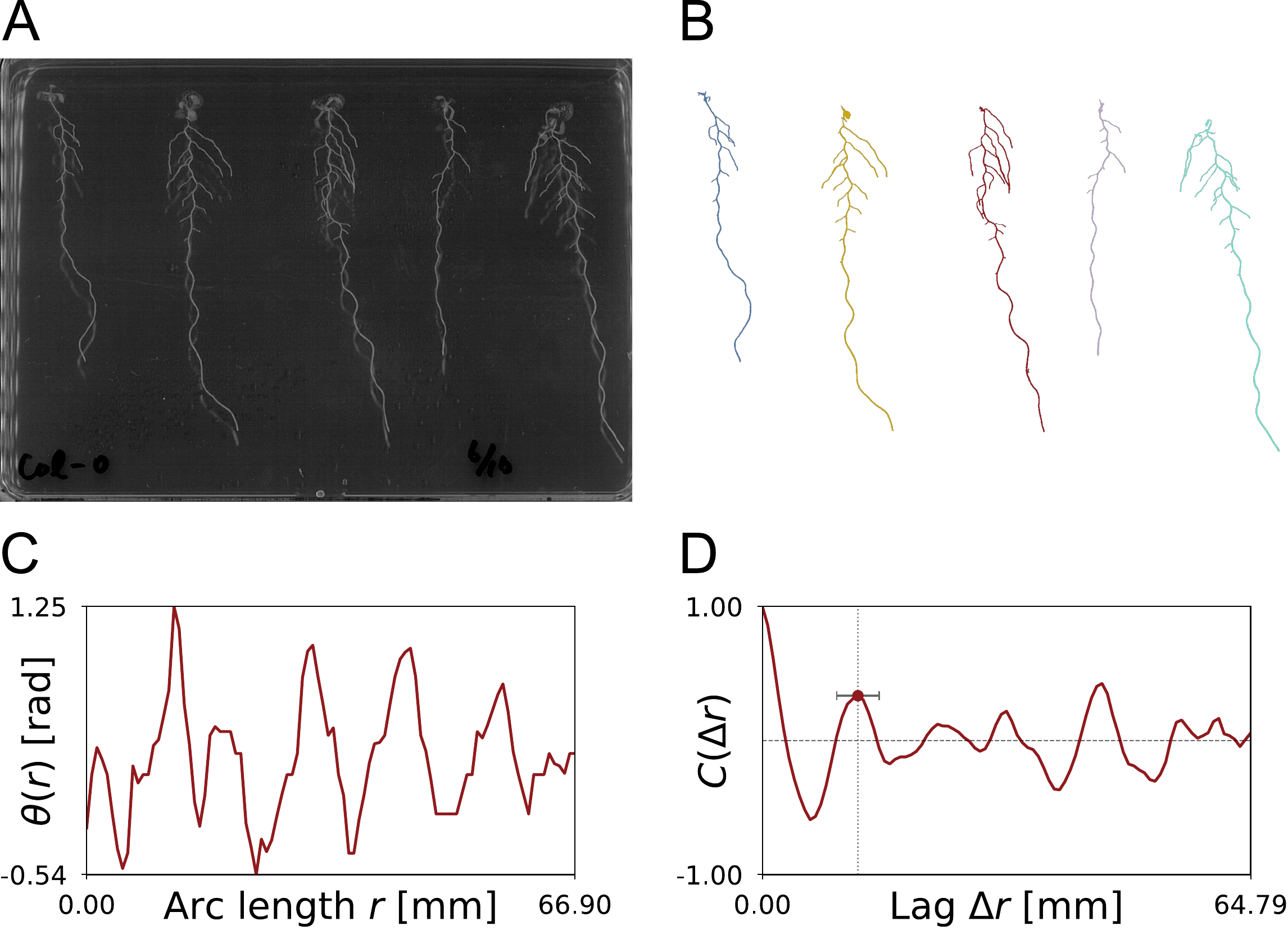}
    \caption{The data analysis pipeline. A: The raw data of \textit{Arabidopsis} growing on an agar plate; B: The roots after being thresholded to create a mask, cleaned and segmented; C: The theta 
    trajectory generated from the middle root in B (after skeletonizing and identifying the primary trunk); D: The unbiased autocorrelation generated from the trajectory. The first peak is the estimate of the arc length traversed in one time period ($\Omega$) while it's width (the error bar) is an estimate for the error. 
    }
    \label{fig:data-schematic}
\end{figure}

First, we convert the raw images (as in \autoref{fig:data-schematic}A) into a single intensity channel, by averaging across the RGB inputs. We then create a binary mask of the root, separating it from the background using thresholding based on pixel intensities. The highlighted pixels are then grouped into connected-component clusters (with the scikit-image package in python \cite{van2014scikit}), and we select only the large clusters to be the `segmented roots' so as to remove background noise (\autoref{fig:data-schematic}B). The segmented roots are then skeletonized with the scikit-image \cite{van2014scikit}. We choose the topmost skeletonized pixel as the `start' node, then find the pixel on the same root at the maximum shortest-path distance from the start node, and define the shortest-path line connecting the two as the primary root (thus separating it from branched lateral roots). We only use the primary root for the analysis, and sample the skeletonized path along it into segments of length 8-pixels. We can then calculate the orientation by calculating the `angle' between the line connecting two consecutive discretized points and the vertical.

This process creates $\theta(r)$, the angle the root makes with the vertical as a function of the distance along the root $r$ (\autoref{fig:data-schematic}C). To obtain an estimate of the arc length traversed in one oscillation period, $\Omega$, we compute the autocorrelation of the zero-mean 
(\autoref{fig:data-schematic}D). After subtracting the mean from $\theta(r)$, we calculate the correlation of each point with itself, and normalize the autocorrelation at each lag $\Delta r$ by the number of points on the root at that lag (as we have several samples for small $\Delta r$, but much fewer for large $\Delta r$). A strongly positive autocorrelation at a lag $\Delta r$ implies that the system at $0$ and $\Delta r$ are correlated, and thus oscillatory autocorrelations indicate the presence of oscillatory behaviour. The arc lengths ($\Omega$) are obtained from the location of the first positive peak, and an estimate of the error in measurement is made through the width of the peak (\autoref{fig:data-schematic}D).

\begin{figure}
    \centering
    \includegraphics[width=\linewidth]{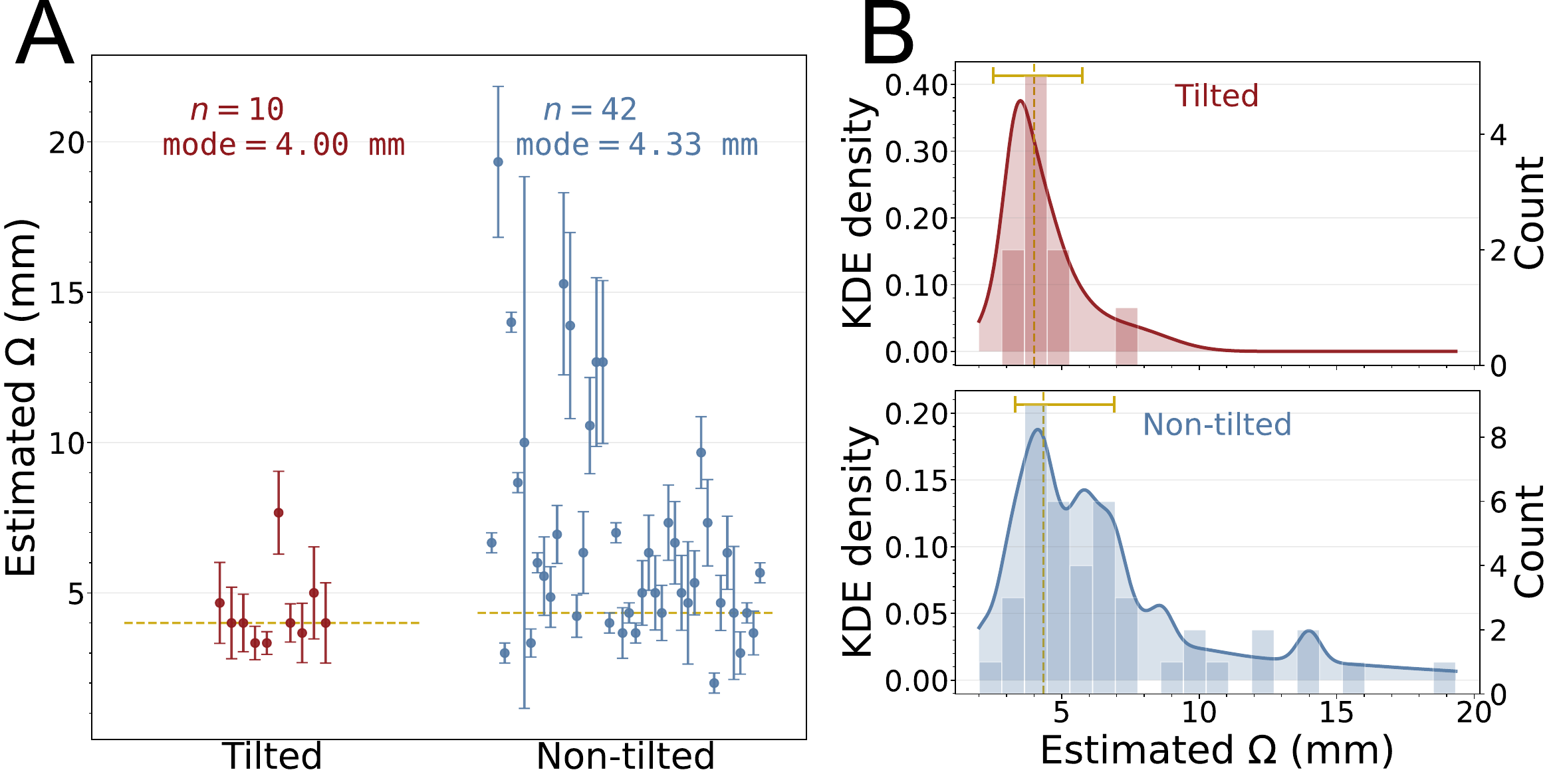}
    \caption{A: Arc lengths $\Omega$ of the oscillations and the error bars, estimated from the first peak of the autocorrelation functions (\autoref{fig:data-schematic}D), stratified by experiments performed on inclined agar plates (red) and vertical ones (blue). Inclined plates show much more robust oscillations, yet experiments on vertical plates also appear to share a similar half-sample mode, as seen by the dashed yellow line. B: The probability distributions of having a certain arc length, for inclined and vertical experiments. The curved line represents the density estimated using Gaussian kernel density estimation (KDE), where each point is taken to be a Gaussian with a width corresponding to the error, which is summed to give a distribution. 
    The bars represent a histogram showing the counts, using the right y-axis. The yellow vertical line represents the mode, while the horizontal bar represents the 95\% confidence interval computed by bootstrap (\autoref{fig:omega-bootstrap}).}
    \label{fig:arclengths-estimation}
\end{figure}

The estimated arc lengths and their distributions, filtered by growth on an incline or vertical, are presented in \autoref{fig:arclengths-estimation}. Interestingly, despite the fact that the oscillations are much clearer on inclined planes (indicated by the coherent data with small errors), several data from the vertical growth position are also observed to have similar arc lengths. While the vertical growth contains a longer tail, several of the larger estimates are imprecise (as can be seen by the error bars in \autoref{fig:arclengths-estimation}A). We estimate the half-sample mode of the distributions, and perform bootstrap resampling to quantify its uncertainity (see \autoref{fig:omega-bootstrap}). For plants grown on vertical agar plates, $\Omega$ is approximated to be $4.33$ mm (while the inclined $\Omega$ is $4.00$ mm).

In a single time period, $T$, the arc length travelled $\Omega$ can be calculated using the root growth speed $v_g$, and further represented by $\tau$ via substituting \autoref{factor-4-equation}:
\begin{align}
    \Omega &= T\, v_g \\
    &= 4\,\tau \, v_g \label{eq:omega-tau-relationship}
\end{align}

The root growth speed of \textit{Arabidopsis} Col-0 under constant illumination, estimated from Beemster and Baskin \cite{beemster1998analysis}, is $\approx 0.241 \pm0.006$ mm/h on the 6th day, and $\approx0.445\pm0.012$ mm/h. We adopt both to generate a rough estimate. With $\tau\approx 2$h \cite{band2012root, baldwin2013gravity}, we get an arc length, $\Omega \approx 1.93 - 3.56$ mm, which partially overlaps with the confidence interval of the directly measured $\Omega$ ($2.51-5.76$ mm for tilted and $3.30 - 6.91$ mm for non-tilted conditions shown by the horizontal bar in \autoref{fig:arclengths-estimation}B; \autoref{fig:omega-bootstrap}). Given the difficulties in measuring $\tau$, our estimate supports \autoref{eq:omega-tau-relationship}.

In the analysis above, we assume a constant root growth speed, and a constant arc length, (as the autocorrelation is, in effect, taking an average across the entire root). In reality, the root growth speed of \textit{Arabidopsis} increases with time \cite{yazdanbakhsh2010analysis, beemster1998analysis, fisahn2012arabidopsis}, with newly germinated roots growing slower than those of week-old plants. Assuming the time delay $\tau$ remains unchanged, under our hypothesis, the arc length, should increase as the plant grows, which qualitatively matches the data: arc lengths near the germ appear shorter than those far away (\autoref{fig:data-schematic}B).

\section{Proposed Experimental Validation}

Further tests could be performed not simply for \textit{Arabidopsis}, but for any oscillating plant root. The time delay could be estimated by tracking auxin (or an auxin indicator like DII venus) levels after a gravitropic stimulus is applied \cite{band2012root}. The time delay could also be estimated by microscopically imaging the cells in the elongation zone after the stimulus: cells on the `inside', close to the vertical would slow down or even completely arrest their growth. The timescale at which this effect is at its maximum after the stimulus would give an estimate of the $\tau$. The time period of oscillations is fairly simple to measure by direct observation, ideally through live imaging. As per our theory, the two should always be approximately a factor $4$ apart.

In addition, the universality of the model across both oscillating and non-oscillating species of plant roots could be experimentally verified. $k$ can be quantified by experiments that measure the gravitropic response strength: the standard practice is to rotate a plant horizontally, and measure the time needed for the growth direction to return to the vertical. $\tau$ could be independently measured as described above. The two could then be multiplied to get an effective value of the control parameter $k\,\tau$, and check whether the plant oscillates. As per our hypothesis, when $k\,\tau$ is below a certain threshold value ($\pi/2$, if the units are scaled correctly) the plant should not show coherent oscillations, but rather damped oscillations and random noise. However, plants which have slower response times or stronger gravitropic responses so that their $k\,\tau$ is above the threshold should show oscillatory dynamics.

\section{Discussion}

We hypothesize that a majority of species will have their parameters for $k$ (the strength of the gravitropic response) and $\tau$ (the time delay, linked to the elongation zone length) that lie in the vicinity of the Hopf bifurcation point. Small changes to the environmental conditions can then create large changes in root dynamics (for example, nutrient rich media may cause oscillations, while nutrient poor media may lead to non-oscillatory growth).

The model's simplicity abstracts from and ignores several phenomena. For example, describing the root's trajectory using just the tip implies that the root growth falls under the `Follow the leader' category described by Thompson and Holbrook \cite{thompson2004root}. However, as they observe, the behaviour is unrealistic as the cell division zone and elongation zone are free to move and readjust until they are locked in place by root hairs. In addition, we neglect to consider the mechanical behaviour which can lead to more complicated phenomena \cite{porat2024quantitative, porat2024mechanical, agostinelli2020nutations}. Nevertheless, the simplicity and generalizability of the model allows it to be applied across several species, and make falsifiable predictions.

The presence of oscillatory behaviour could also be beneficial from an evolutionary perspective, as it helps a system explore its environment. In a simple case of periodic obstacles the problem resembles a Frenkel-Kontorova model \cite{frenkel1938theory}: if the periodicity is incommensurate with that of the natural frequency of the oscillations, chaotic dynamics could be observed \cite{aubry1983twist, aubry1984trajectories}. These chaotic dynamics could help a root explore its environment in an efficient way. While fractal-like patterns of root systems are largely attributed to branching, chaotic dynamics within each individual root could also be a factor. The deterministic system can show chaotic behaviour even without obstacles, for large values of $k\,\tau$, yet introducing noise can significantly lower the threshold, leading to chaotic behaviour even at morderate values (\autoref{fig:chaos-noise-boundary}).

Intraspecific variations between different ecotypes (natural accessions) is of particular relevance due to the possibility of phenocopies \cite{goldschmidt1949phenocopies, waddington1953genetic}. Experiments show that external perturbations can induce phenotypic changes in a species, the likes of which are seen in mutants of the same species. In the same vein, perhaps the same wild-type root can show growth behavior resembling that of mutant strains depending on conditions in its environment, such as the presence of rocks in soil, water and nutrient concentration, etc. Ammonium concentration, for instance, is known to effect gravitropism \cite{zou2012ammonium} by delaying the onset of an asymmetrical auxin gradient, indicating that not just $\tau$, but also $k$ can be changed due to environmental factors. In extreme conditions, it could be possible that drastic concentrations cause chaotic behavior by causing a large increase in the control parameter $k\,\tau$.

In addition, this model is best to represent oversimplified laboratory conditions. Nutrient concentrations in real soil vary \cite{sardans2017plant}, and plant roots are known to chemotax or even branch towards them \cite{izzo2019chemotropic}. In addition, soil particles could be present as obstacles hindering growth, or stable surfaces to grip upon. The plant roots could simultaneously affect and be effected by the environment in a feedback loop \cite{noronha2025modeling, van2013plant}.

In conclusion, we develop a simple stochastic delay differential equation, mainly controlled by a single parameter $k\,\tau$, a product of the gravitropic response strength and the time delay in the response. Depending on the value of this parameter, a plant root undergoes a delay-induced Hopf bifurcation, going from vertical growth to oscillatory behaviour with period 4$\tau$, which appears universal across many non-linear responses. By adding noise, we can observe a transition to chaos through intermittency, allowing for chaotic behaviour even at low $k\,\tau$, which could be beneficial to explore an environment. The arc length, predicted using experimental data on the time delay $\tau$ appears to reasonably match arc lengths measured from experiments. The simplicity of our theory allows it to be universally applied to several root species, allowing for easy prediction and validation.

\section{Acknowledgements}

We thank Tatsuaki Goh for kindly providing {\it Arabidopsis} root pictures and Tohya Suzuki for helpful discussions. RFN and KK are supported by the Novo Nordisk Foundation Grant No. NNF21OC0065542. KF is supported by the Japan Society for the Promotion of Science Grant No. 26H00457.

\printbibliography

\appendix

\section{Analytical calculations for the linear model} \label{appendix:linear-model}

The delay differential equation is 
$$ \frac{\textrm{d}\theta}{\textrm{d}t} = - k \,\sin(\theta(t-\tau))$$
For small $\theta$, we can assume $\sin\theta\approx\theta$, converting it to a linear DDE:
$$ \frac{\textrm{d}\theta}{\textrm{d}t} = - k \,\theta(t-\tau)$$
Make the ansatz
$$ \theta(t) = A\, e^{\lambda t} $$
Then, 
\begin{align*}
\frac{\textrm{d}\theta}{\textrm{d}t} &= A\lambda\, e^{\lambda t} \\
\theta(t-\tau) &= A\,e^{\lambda(t-\tau)} \\
&= A\,e^{\lambda t} \,e^{-\lambda \tau}
\end{align*}
Substituting in,
\begin{align*}
\frac{\textrm{d}\theta}{\textrm{d}t} &= - k \,\theta(t-\tau) \\
A\,\lambda\, e^{\lambda t} &= -k \, A\,e^{\lambda t} \,e^{-\lambda \tau} \\
\lambda &= -k\,e^{-\lambda \tau}
\end{align*}
Which gives us the characteristic equation
$$\lambda + k\,e^{-\lambda \tau} = 0$$
This is a transcendental equation, that needs to be solved for $\lambda$. $\lambda$ is, in principle, complex, and thus it can be written as 
$$ \lambda = a + bi$$
where $a$ controls the amplitude of the oscillations. If $a>0$, the oscillations amplify, while if $a<0$, they're damped. Thus, the critical boundary can be found to be where $a=0$:
\begin{align*}
b\,i + k\,e^{-b\,i\, \tau} &= 0 \\
b\,i + k\,(\cos(b\,\tau) - i \sin(b\,\tau)) &= 0
\end{align*}
The real part gives us:
$$ k \, \cos(b\,\tau)=0 $$
Which implies that
$$b\tau = \frac\pi2 + n\pi ;\qquad n=\{0,1,2,3,..\}$$
The imaginary part gives us
$$
b - k\, \sin(b\,\tau)=0$$
$$
b = k
$$
Thus, the critical boundary occurs at 
$$ k \cdot\tau = \frac\pi2 $$
Note that $b$ represents the angular velocity ($\omega$). The time period can thus be written as:
\begin{align*}
T &= \frac{2\,\pi}{b} \\
&= \frac{2\,\pi}{k} \\
&= 2\,\pi \left(\frac{2\,\tau}{\pi}\right) \\
&= 4\, \tau
\end{align*}

\section{Amplitude of the first limit cycle} \label{appendix:amplitude-scaling}

Assume a sinusoidal solution, of the type:
$$\theta(t) = R\cos(\omega t)$$
Then, 
\begin{align*}
\frac{d\theta}{dt} &= -R\omega \sin(\omega t) \\
\theta(t-\tau) &= R\cos(\omega t - \omega\tau)
\end{align*}
Our DDE is 
$$ \frac{d\theta}{dt} = -k \,\sin(\theta(t-\tau)) $$
We need to consider $\sin(R\cos(\phi))$, as that appears in the RHS of the DDE. That can be expanded using the real-valued Jacobi–Anger expansion:
$$\sin(R\cos\phi) = 2\sum_{n=0}^{\infty}(-1)^n J_{2n+1}(R)\cos((2n+1)\phi)$$
Where $J_n(x)$ is the $n$-th Bessel function of the first kind.
Keeping only the fundamental ($n=0$):
\begin{align*}
\sin(R\cos\phi) &\approx 2 J_1(R)\cos(\phi) \\
-k\sin(\theta(t-\tau)) &\approx -2k J_1(R)\cos(\omega t - \omega\tau)
\end{align*}
Now, matching that to the LHS of the DDE, $d\theta/dt$, we get
\begin{align*}
-R\omega \sin(\omega t) &= -2k J_1(R)\cos(\omega t - \omega\tau) \\
R\omega \sin(\omega t) &= 2k J_1(R)\cos(\omega t - \omega\tau) \\
\end{align*}
Using $\cos(x-y) = \cos(x)\cos(y)+\sin(x)\sin(y)$,
\begin{align*}
R\omega \sin(\omega t) &= 2k J_1(R) \big[\cos(\omega t)\cos(\omega\tau) + \sin(\omega t)\sin(\omega\tau) \big] \\
\end{align*}
Matching coefficients of $\cos(\omega t)$:
$$ 
2kJ_1(R)\cos(\omega\tau) = 0 
$$
$$
\omega\tau = \pi/2
$$
This gives us $w=\pi/2\tau$: Omega is fixed by $\tau$ alone (independent of $k$) and furthermore, the time period $T=2\pi/\omega = 4\tau$ (confirming what we expected).

Next, matching the $\sin(\omega t)$ coefficients gives us:
\begin{align*}
R\omega &= 2kJ_1(R)\sin(\omega\tau) \\
&= 2kJ_1(R)
\end{align*}
With $\omega = \pi/(2\tau)$:
\begin{align*}
R\,\frac{\pi}{2\tau} &= 2kJ_1(R)\sin\left(\frac{\pi\tau}{2\tau}\right) \\
\frac{J_1(R)}R &= \frac\pi{4k\tau}
\end{align*}
$$\frac{2J_1(R)}{R} = \frac{\pi}{2k\tau}$$
Rearranging this equation, given the control parameter $k\tau$, we can calculate the amplitude $R$ to be
$$k\tau = \frac{\pi R}{4J_1(R)}$$

\section{Characterising the bifurcations of the second limit cycle} \label{appendix:ghost-escape-times}

\begin{figure}[h]
    \centering
    \includegraphics[width=\linewidth]{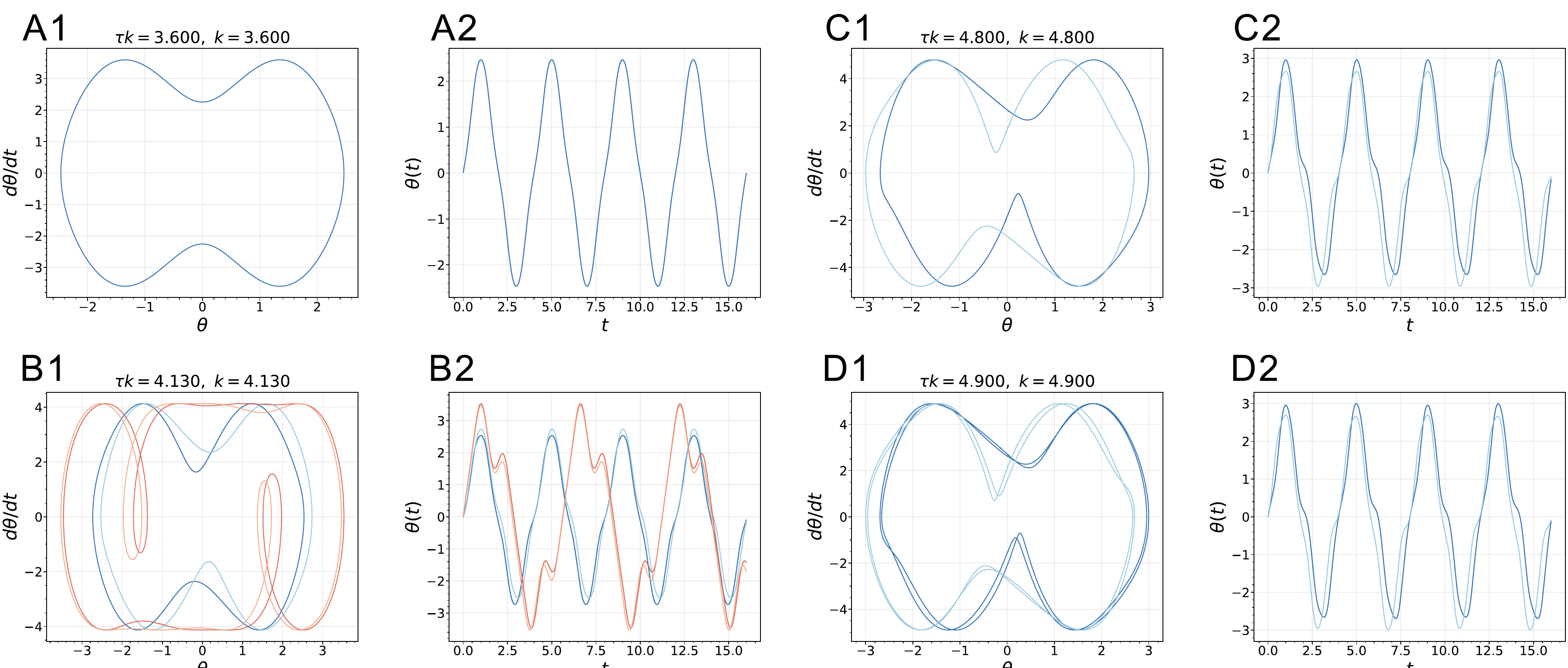}
    \caption{Phase portraits (1) of the limit cycles (timeseries shown in 2) observed for different parameters: A: $\tau k=3.6$, before the point symmetry breaking; B: $\tau k=4.13$, showing the two symmetric limit cycles of period 4 (blue) as well as the second limit cycle (orange) that emerges and quickly splits into two; C: $\tau k=4.8$, after the second limit cycle disappears, the system only contains the two period-4 limit cycles (similar to parameters after A before the second limit cycle appears); D: $\tau k=4.9$, the final bifurcation before the system approaches chaos through intermittency, in the form on a period-doubling. Only in this narrow region near the chaotic boundary are limit cycles of period 4 not observed in the system. It is also important to stress that this is not a period doubling cascade, but a single event before intermittency.}
    \label{fig:timeseries-phaseplots}
\end{figure}

In order to analyze the bifurcations that give rise to the second limit cycle (which can be seen in \autoref{fig:timeseries-phaseplots}B), we investigate the ghost escape times. The second limit cycle is only stable in a narrow interval, and so we can start a simulation within that interval, and obtain the steady state limit cycle. Then, using that as the initial condition, we can change the parameter $k\,\tau$ to where the limit cycle ceases to exist, and track how long it takes for the system to relax onto the original period 4 limit cycle. Close to the bifurcation boundary, it is possible for the system to be stuck in the ghost orbits for several time periods.

\begin{figure}[h]
    \centering
    \includegraphics[width=\linewidth]{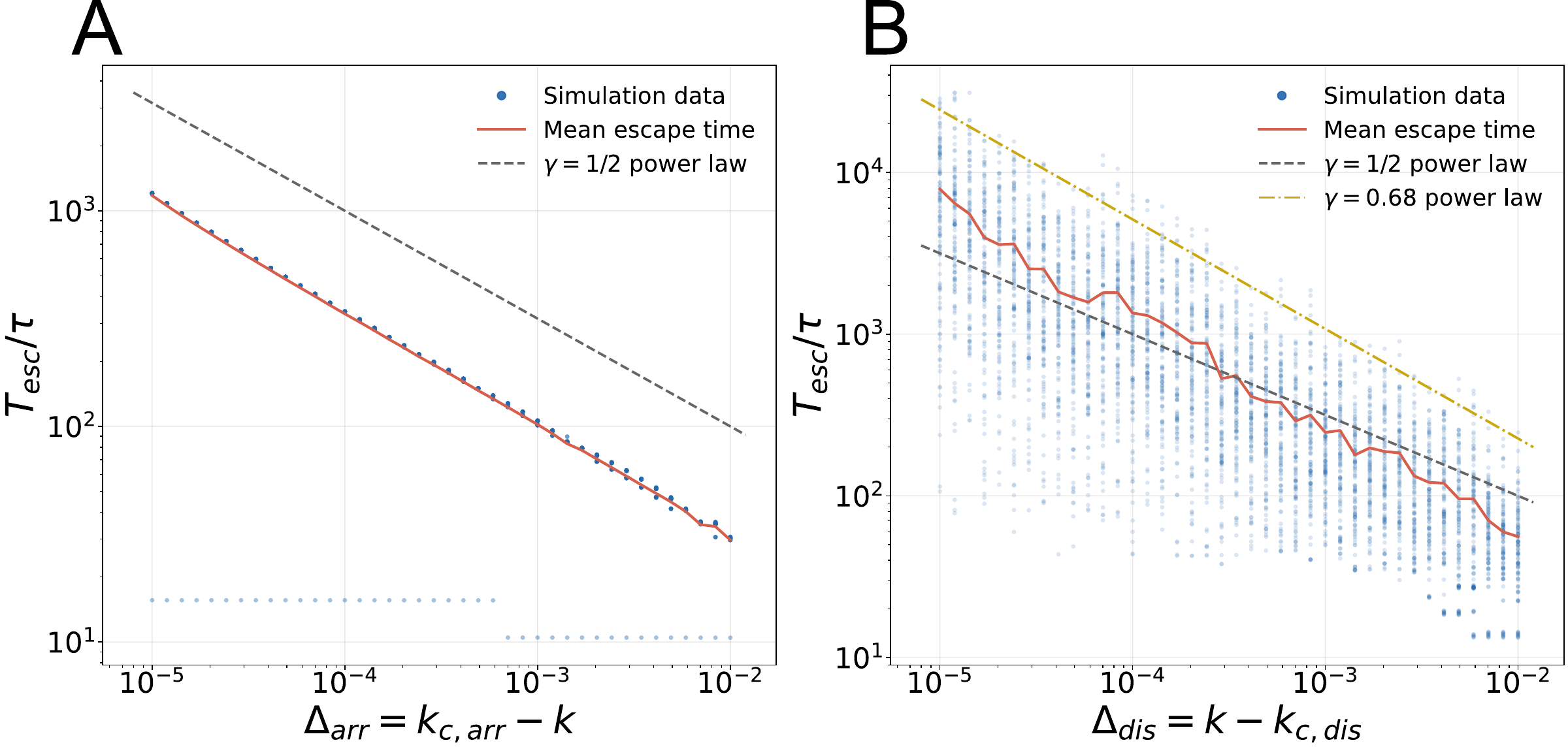}
    \caption{Ghost escape times for the second limit cycle. A: Onset of the second limit cycle: a power law of $t\sim\Delta^{-\frac12}$ is clearly visible, signifying a saddle-node bifurcation of limit cycles. Note that the previously described condition of having a single point in the initial function outside $[-\pi,\pi]$ is not strong enough, as a few points escape quickly. B: Disappearance of the limit cycle. Note that the points are much more scattered, and the power law of $\approx 0.68$ is distinct from $0.5$, signifying a crisis.}
    \label{fig:lc2-ghost-escape-times}
\end{figure}

We perform simulations for several different parameter values close to the critical point, for $\tau=1$. In addition, we simulate for $120$ initial functions, sequentially sampled by shifting along the limit cycle. The second limit cycle has period $\approx 5.5\tau$, and our initial function is of length $\tau$: if we choose initial functions close to 0, the system will instantly go to first (always stable) limit cycle of period 4, regardless of how close we are to the bifurcation. As such, we only choose initial conditions where at least one point in the initial function has a value outside the interval $[-\pi,\pi]$. The results are visualized in \autoref{fig:lc2-ghost-escape-times}.

\section{Effect of noise on the boundary to chaos} \label{appendix:lyapunov-noise}

We numerically measure the largest Lyapunov exponent of the system, for varying values of noise strength $\eta$ and the control parameter $\tau k$. After warming up the system for $500\tau$ timesteps this corresponds to $\approx125$ time periods), we create a small perturbation $\delta_0$ to the entire history of the function. We then evolve both trajectories with identical noise: the `random numbers' drawn at each timestep are the same, so that the divergence primarily comes from dynamics, and not a difference in random numbers.

\begin{figure}[h]
    \centering
    \includegraphics[width=\linewidth]{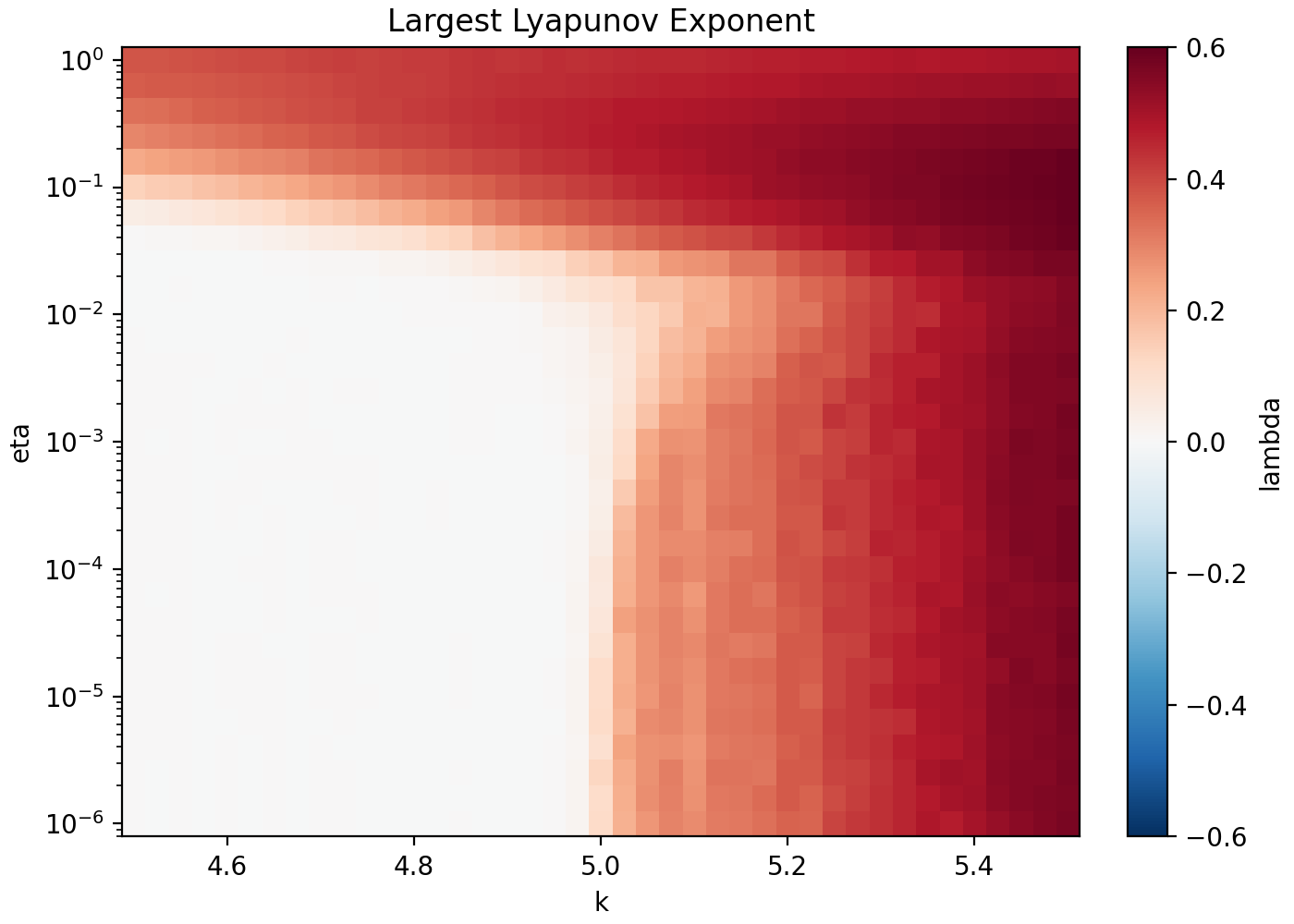}
    \caption{A heatmap of Lyapunov exponent values against noise strength $\eta$ and control parameter $\tau k$. For low noise strengths, the boundary to chaos appears relatively constant regardless of the noise. However, as the noise increases, we observe that noise causes the boundary of chaos to occur sooner. This can be easily understood, as the chaos appears through intermittency: the domain is divided into intervals of length $2\pi$, and chaos is characterized by the attractor staying in one interval for some time, before a quick jump to another. In this way, noise could help a laminar }
    \label{fig:chaos-noise-boundary}
\end{figure}

Every $\tau$ timesteps, we compare the entire history buffer (as the system's state for a delay-differential equation consists not just the current value $\theta(t)$, but rather the entire history $\theta(t^\prime), \forall \, t-\tau \leq t^\prime \leq t$). We then compute the RMS distance, $d$, by
$$d = \sqrt{\frac1N\sum_{i=1}^{N}(\theta_i^{\text{pert}}-\theta_i^{\text{base}})^2}$$
After this, the deviation on the entire history buffer is renormalized to be $\delta_0$. We then repeat this process for $N_\text{renorm} = 2000\tau$ timesteps, generating multiple $d_j$ values. The largest Lyapunov exponent is thus calculated by
$$\lambda \approx \frac{1}{N_{\text{renorm}}}
\sum_{j=1}^{N_{\text{renorm}}}\log\!\left(\frac{d_j}{\delta_0}\right)$$
Due to the stochasticity, we perform the same algorithm for 32 runs, and report the results in \autoref{fig:chaos-noise-boundary}.

Low noise values do not significantly change the boundary to chaos, but once the noise gets large at around $\eta=0.1$, the boundary shifts fairly quickly. This can allow for chaotic growth even at moderate values of the gravitropic response strength $k$ and time lag $\tau$, which could be evolutionarily beneficial and could exist in certain species of roots.

\section{Mode error estimation from data}

\begin{figure}[h]
    \centering
    \includegraphics[width=\linewidth]{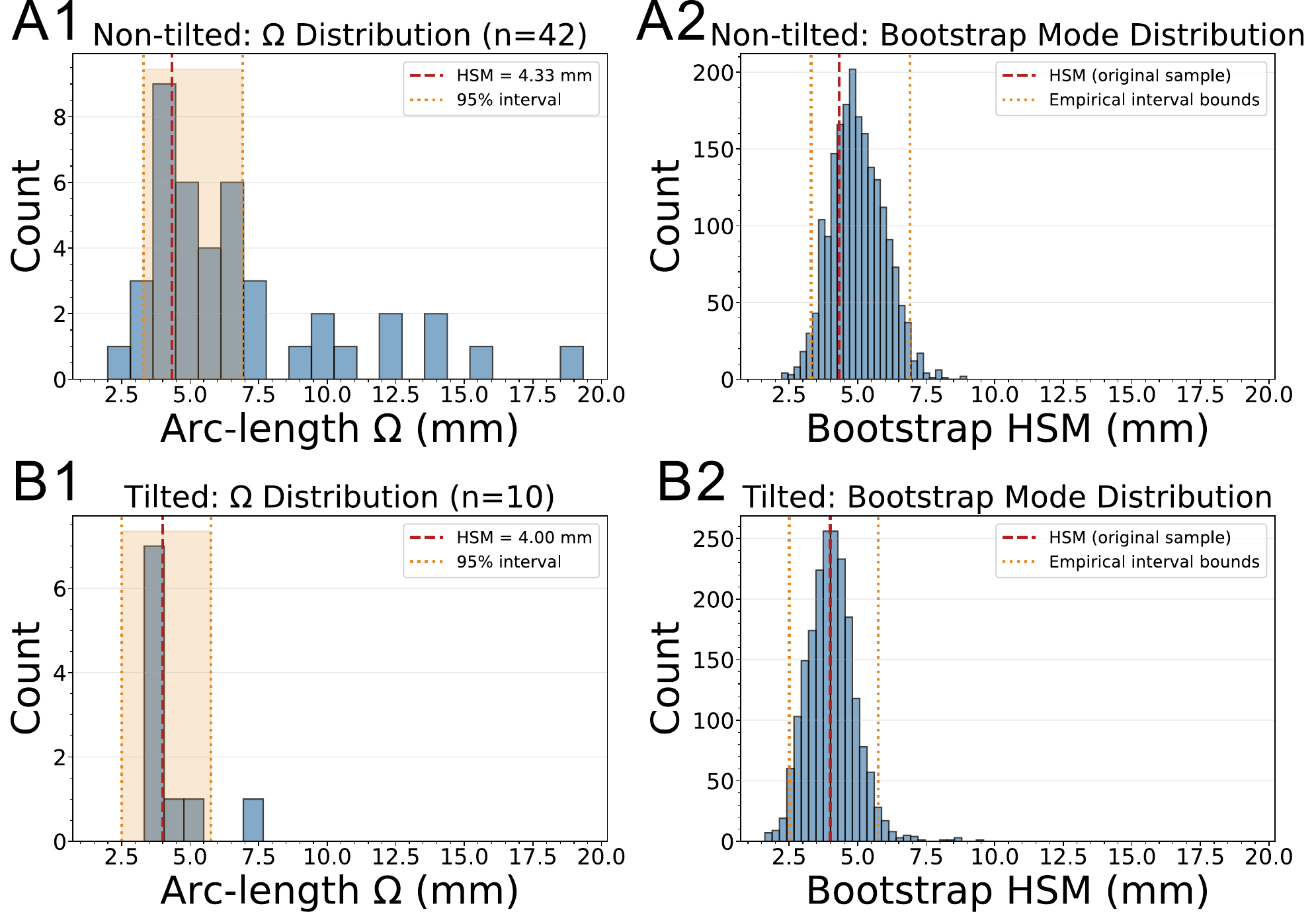}
    \caption{The mode is generated through the half-sample mode of the data. A1: The distribution of the non-tilted data, as in \autoref{fig:arclengths-estimation}, with the mode and 95\% confidence interval labelled. A2: The 2000 bootstrapped samples, used to create the confidence intervals. Samples are smoothed by Gaussian noise derived from their standard deviations. The 95\% CI is found to be $[3.30, 6.51]$ B1, B2: The same analysis done on the roots grown on inclined agar plates, with a CI of $[2.51, 5.76]$.}
    \label{fig:omega-bootstrap}
\end{figure}

The half-sample mode is estimated directly from the data, while the confidence intervals are estimated through bootstrapped samples (\autoref{fig:omega-bootstrap}).

\end{document}